# Continuous Collapse of the Spin Cycloid in BiFeO₃ Thin Films under an Applied Magnetic Field probed by Neutron Scattering

Md. Firoz Pervez,[1] Hongrui Zhang,[2,3] Yen-Lin Huang,[2,4] Lucas Caretta,[4,5] Ramamoorthy Ramesh,[2,4,6,7,8,9 †] and Clemens Ulrich[1,*]

[1] School of Physics, The University of New South Wales, NSW 2052, Australia

[2] Materials Science Division, Lawrence Berkeley National Laboratory, Berkeley, CA, 94720, USA.

[3] Ningbo Institute of Materials Technology & Engineering, Chinese Academy of Sciences, Ningbo 315201, China

[4] Department of Materials Science and Engineering, University of California, Berkeley, CA 94720, USA

[5] School of Engineering, Brown University, Providence, RI, 02912, USA.

[6] Department of Physics, University of California, Berkeley, CA, 94720, USA.

[7] Materials Science and NanoEngineering, Rice University, Houston, TX, 77005, USA.

[8] Departments of Chemistry, and Physics and Astronomy, Rice University, Houston, TX, 77005, USA.

[9] Rice Advanced Materials Institute, Rice University, Houston, TX, 77005, USA.

[*] Corresponding Author: c.ulrich@unsw.edu.au

[†] Corresponding Author: rramesh@berkeley.edu

Md. Firoz Pervez, ORCID: 0000-0002-1453-8704

Clemens Ulrich, ORCID: 0000-0002-6829-9374

Hongrui Zhang, ORCID: 0000-0001-7896-019X

Yen-Lin Huang, ORCID: 0000-0002-6129-8547

Lucas Caretta, ORCID: 0000-0001-7229-7980

Ramamoorthy Ramesh, ORCID:  0000-0003-0524-1332

**ABSTRACT**

Bismuth ferrite ($BiFeO_3$) is one of the rare materials that exhibits multiferroic properties already at room-temperature. Therefore, it offers tremendous potential for future technological applications, such as memory and logic. However, a weak magnetoelectric coupling together with the presence of a noncollinear cycloidal spin order restricts various practical applications of $BiFeO_3$. Therefore, there is a large interest in the search for suitable methods for the modulation of the spin cycloid in $BiFeO_3$. By performing neutron diffraction experiments using a triple-axis instrument we have determined that the spin cycloid can be systematically suppressed by applying a high magnetic field of 10 T in a $BiFeO_3$ thin film of about 100 nm grown on a (110)-oriented $SrTiO_3$ substrate. As predicted by previous theoretical calculations, we observed that the required critical magnetic field to suppress the spin cycloid in a $BiFeO_3$ thin film was lower as compared to the previously reported critical magnetic field for bulk $BiFeO_3$ single crystals. Our experiment reveals that the spin cycloid continuously expands with increasing magnetic field before the complete transformation into a G-type antiferromagnetic spin order. Such tuning of the length of the spin cycloid up to a complete suppression offers new functionalities for future technological applications as in spintronics or magnonics.

# 1. INTRODUCTION

Multiferroics are a fascinating class of materials that manifest multiple, simultaneous ferroic orders such as ferroelectric and magnetic polarizations. Magnetoelectric coupling between the ferroelectric and magnetic orders, including ferromagnetic, ferrimagnetic, antiferromagnetic or more complex noncollinear spin structures has been observed in single or multiphase multiferroic materials. Often this magnetoelectric coupling emerges from the simultaneous breaking of spatial inversion symmetry and time-reversal symmetry [1-2]. A magnetoelectric coupling in multiferroics allows for an efficient method of the electric control of the magnetic order as well as the magnetic control of electric polarization. Therefore, the magnetoelectric coupling offers an enormous potential for the application in next-generation spintronics and magnonic devices (e.g., memory devices, magnetic switches, magnetic sensors, high-frequency magnetic devices, spin valve devices, etc.) at much lower power consumption and faster operation as compared to the present conventional electronic devices [1-13].

Bismuth ferrite ($BiFeO_3$) is one of the rare room temperature type-I, single-phase multiferroic oxides where primarily independent magnetic and ferroelectric order emerge [1,9]. However, a weak direct coupling between magnetic and ferroelectric order is present in $BiFeO_3$ [7,14,15]. Bulk $BiFeO_3$ is a rhombohedrally distorted cubic perovskite with the polar space group $R3c - C_{3v}^6$ [16] and a lattice parameter of 3.968 Å in the pseudocubic (pc) notation. This notation will be used throughout this article. In the case of $BiFeO_3$ thin films, epitaxial tensile or compressive strain results in a further distortion of the in-plane and out-of-plane lattice parameters. As consequence, different crystal structures including tetragonal ($P4mm$), monoclinic ($Cm$), monoclinic ($Cc$), rhombohedral ($R3c$) and orthorhombic ($Ima2$) are possible in $BiFeO_3$ thin films [17,18].

In the $BiFeO_3$ single crystals, the displacement of the A-site $Bi^{3+}$ ions and the lone pair $s$-electrons cause the rhombohedral distortion which leads to the spontaneous ferroelectric polarization of about $100 \, \mu C \, cm^{-2}$ along the direction of the body diagonal $[111]_{pc}$. This spontaneous ferroelectric polarization persists up to high temperatures ($T_C \sim 1123$ K) [19,20]. Below 643 K $BiFeO_3$ possesses a complex noncollinear magnetic structure arising from the $Fe^{3+}$ ions along with nearly zero average magnetization [21]. The $Fe^{3+}$ magnetic moments of the nearest neighbors of the adjacent $(111)_{pc}$ planes are ordered in a predominantly G-type antiferromagnetic spin structure. The Dzyaloshinskii-Moriya interaction, which is caused by both, the spin-orbit coupling and a broken inversion symmetry, tends to favor a particular spin rotation in $BiFeO_3$ which leads to a cycloidal spin order with a propagation length of about $\lambda = 63$ nm. In bulk $BiFeO_3$ the spin cycloid propagates along one of three crystallographic directions: $[\bar{1}10]_{pc}$, $[10\bar{1}]_{pc}$ or $[01\bar{1}]_{pc}$ where the rotation axis of the spins is perpendicular to the plane defined by the propagation vector and the direction of the electric polarization along $[111]_{pc}$ [20-25].

Through the existence of s spin cycloid in $BiFeO_3$ thin films was long time under debate, in 2010 Ke *et al.* found a type-I spin cycloid ($\sim 64$ nm ) propagating along the $[1\bar{1}0]_{pc}$ direction in a partially relaxed $BiFeO_3$ thin film with a thickness of about 800 nm deposited on a (100)-oriented $SrTiO_3$ substrate [26]. In 2011, Ratcliff *et al.* discovered the presence of a spin cycloid ($\sim 62$ nm) in a 1 $\mu$m thin $BiFeO_3$ film deposited on a (110)-$SrTiO_3$ substrate [27], propagating along a unique $[11\bar{2}]_{pc}$ direction, which is different from the direction $[1\bar{1}0]_{pc}$ in the bulk single crystals [24,28]. In a neutron scattering study, Bertinshaw *et al.* discovered a spin cycloid in a $BiFeO_3$ thin film of just 100 nm when grown on a (110)-oriented $SrTiO_3$ substrate with a thin $SrRuO_3$ intermediate layer [29]. This report also showed that the length of the spin cycloid extends and diverges to infinity at the magnetic phase transition temperature of $650 \pm 10$ K. Finally, the report of Burns *et al.* revealed a dependence of the length of the spin cycloid on the film thickness. They observed that the length of spin cycloid increases for decreasing film thicknesses down to 20 nm [30].

Various factors can influence the spin cycloid in $BiFeO_3$ thin films such as temperature, external electric and magnetic fields, film thickness, epitaxial strain, an intermediate layer, doping, etc. [8,10,26-34]. One of the key experimental strategies for understanding the fundamental physical interactions and domain states in magnetic materials is applying an external magnetic field. Several experimental techniques, including magnetization measurements and electric polarization measurements [7,14,15,35-38], electron spin resonance experiments (ESR) [39] and Raman light scattering experiments [40] have been performed to investigate the influence of an applied magnetic field (pulsed and static) on $BiFeO_3$. They showed that applying a high magnetic field can suppress spin cycloid in $BiFeO_3$ single crystals or polycrystalline ceramics at a magnetic field of 16 T to 20 T [7,14,15,35-38,41,42]. A theoretical study by Gareeva *et al.* predicted the lowering of the required critical magnetic field for strained $BiFeO_3$ thin films as compared to $BiFeO_3$ single crystals [43,44]. Moreover, the Raman experiments by Agbelele *et al.* on $BiFeO_3$ thin films [40] reported a magnetic phase transition from a cycloidal spin structure to a G-type antiferromagnetic spin order in the range of 4 T−6 T for BiFeO$_3$ thin films grown on different substrates.

Neutron scattering experiments would unambiguously reveal the nature of magnetic phase transition and would determine the precise magnetic spin structure in BiFeO$_3$ thin films. Therefore, we have performed neutron diffraction experiments on a 100 nm thick BiFeO$_3$ film deposited by the PLD technique on a (110)-oriented SrTiO$_3$ substrate. The experiment was performed using a triple-axis spectrometer at high magnetic fields of up to 10 T, in order to investigate the overall effect of a magnetic field on the spin structure in BiFeO$_3$ thin films.

## 2. EXPERIMENTAL DETAILS

The BiFeO$_3$ thin film sample with a film thickness of ∼100 nm was grown by the Pulsed Laser Deposition on a (110)-oriented single-sided polished, 10 mm × 10 mm × 0.5 mm SrTiO$_3$ substrate (SHINKOSHA CO., LTD.). As an SrRuO$_3$ intermediate layers can become magnetic at low temperature, the sample was grown without an intermediate layer. The lattice mismatch between SrTiO$_3$ ($a_{pc}$ = 3.905 Å [45]) and BiFeO$_3$ ($a_{pc}$= 3.968 Å [16]) leads to an out-of-plane elongated lattice parameter, resulting in a monoclinically distorted structure of the BiFeO$_3$ films [18,27,29,30,46,47].

The triple-axis spectrometer instrument TAIPAN [48], located at the Australian Centre for Neutron Scattering (ACNS) at the Australian Nuclear Science and Technology Organisation (ANSTO) in Sydney, Australia, has demonstrated to be the ideal choice for the measurement of magnetic Bragg peaks of transition metal oxide thin films due to its enhanced resolution and the improved signal-to-background ratio as compared to conventional neutron diffraction instruments [29,30,49]. In order to apply a high magnetic field, a 12 T superconducting magnet was used. The magnetic field was applied in the direction perpendicular to the scattering plane, i.e. along the $[1\bar{1}0]$ in-plane direction of the film. In order to suppress contaminations from second order reflections from the SrTiO$_3$ substrate at the $\left(\frac{1}{2}\ \frac{1}{2}\ \frac{1}{2}\right)$ Bragg peak position, two pyrolytic-graphite, PG (002), filters with a total thickness of 60 mm were placed behind the sample. An incident energy of 14.86 eV ($\lambda = 2.3462$ Å) was used in the standard elastic scattering mode. This was provided by a PG-monochromator and a PG-analyzer with vertical focusing but horizontally flat configuration. In order to enhance the resolution and to suppress the background, $40''$ collimators were placed in the neutron beam before and after the sample (see also Ref. [29,30,49]). A linescan was performed around the $\left(\frac{1}{2}\ \frac{1}{2}\ \frac{1}{2}\right)$ Bragg peak of the second order contamination of the $SrTiO_3$ substrate without the two PG (002) filters to determine the instrumental resolution of a Full Width at Half Maximum of 0.0032 reciprocal lattice units (r.l.u.).

The substrate $SrTiO_3$ possesses a crystallographic phase transition at 105 K. In order to avoid corresponding effects on the $BiFeO_3$ thin film, the neutron scattering measurements were performed at the temperature of 150 K [49], which also provides a stronger magnetic signal from the $BiFeO_3$ thin film compared to room temperature [29].

## 3. RESULTS AND DISCUSSION

Figure 1 shows the neutron diffraction data of the 100 nm thick $BiFeO_3$ film taken at a temperature of 150 K and zero magnetic field. In order to access the magnetic Bragg peaks arising from the spin cycloid, the film was grown on a (110)-oriented $SrTiO_3$ substrate and mounted in the [110] x [001] scattering plane. Figure 1(a) shows the neutron scattering reciprocal space map (RSM) around the $\left(\frac{1}{2}\ \frac{1}{2}\ \frac{1}{2}\right)_{pc}$ Bragg reflection. The peak at precisely $\left(\frac{1}{2}\ \frac{1}{2}\ \frac{1}{2}\right)_{pc}$ arises from the second order contamination of the (111) structural Bragg reflection of the $SrTiO_3$ substrate and is still visible despite the use of two PG (002) filters. The two Bragg peaks at around $Q_{HK}$ = 0.485 can be attributed to magnetic scattering from the spin cycloid of the $BiFeO_3$ thin film. The reduced $Q_{HK}$ value is a consequence of the epitaxial strain, which leads to an increase of the out-of-plane lattice parameter [29,30]. The distance between the peak maxima indicates the length of the spin cycloid ($\lambda = 2\pi/\delta$) and the direction of their separation, as shown by the black line, corresponds to the propagation direction of the spin cycloid along $[11\bar{2}]_{pc}$. This propagation direction was also observed by previous neutron diffraction experiments on $BiFeO_3$ thin films [26,27,29,30]. Theoretical studies have predicted that a spin cycloid can possibly propagate either along the $[11\bar{2}]_{pc}$, $[1\bar{2}1]_{pc}$ or $[\bar{2}11]_{pc}$ directions in a $BiFeO_3$ thin film deposited on a (110)-oriented $SrTiO_3$ substrate as compared to the $[\bar{1}10]_{pc}$, $[10\bar{1}]_{pc}$ or $[01\bar{1}]_{pc}$ directions in bulk $BiFeO_3$ [27,50]. The change in the propagation direction is a consequence of the in-plane strain of the $BiFeO_3$ film.

Figure 1(b) shows the RSM around the $\left(\frac{1}{2} \frac{1}{2} \frac{\bar{1}}{2}\right)_{\text{pc}}$. Only one magnetic Bragg peak appears. Due to the orientation of the instrumental resolution ellipsoid the two magnetic Bragg peaks of the spin cycloid are overlapping in this measurement geometry (see Suppl. Mater of Ref. [30]). Figures 1(c) and 1(d) show the corresponding line scans along the black lines in Figs. 1(a) and (b), respectively. The red lines are Gaussian lineshapes fitted to the experimental data. For the fit the linewidth was kept constant for both magnetic Bragg peaks. The separation of the two magnetic Bragg peaks in Fig. 1(c) is $\delta = \Delta Q_H = \Delta Q_K = \Delta Q_{HK} = 0.00415$, which corresponds to a length of the spin cycloid of $65.7 \pm 1.0$ nm. This is in good agreement to the value of about 63 nm obtained for bulk $BiFeO_3$ [20-25] and for $BiFeO_3$ thin films [27,29,30]. Figure 1(d) shows the corresponding linescan over the $\left(\frac{1}{2} \frac{1}{2} \frac{\bar{1}}{2}\right)_{\text{pc}}$ RSM as indicated by the black line in Fig. 1(b). Note, to avoid the presence of the $\lambda/2$ contamination from the $SrTiO_3$ substrate, the linescan was not taken precisely through the peak maximum but slightly shifted as indicated by the black line in Fig. 1(b). Only one sharp peak is observed. Its integrated intensity corresponds roughly to the integrated intensity of the two magnetic Bragg peaks in Fig. 1(b) [27,29,30].

The magnetic field dependent data are shown in Fig. 2. The data were taken around the $\left(\frac{1}{2} \frac{1}{2} \frac{1}{2}\right)_{\text{pc}}$ Bragg reflection following the black line of the RSM as shown in Fig. 1(a), i.e. along the $[11\bar{2}]_{\text{pc}}$ direction. The magnetic field was increased in 1 T steps to a strength of 10 T (Fig. 2(a)) and then decreased to zero field again (Fig. 2(b)). The obtained data were analyzed by fitting two Gaussian lineshapes to the experimental data, where the linewidth and peak intensity of each of the two Gaussian peaks were kept constant using the values obtained for the 0 T measurement. With increasing magnetic field the distance between the two magnetic Bragg peaks decreases, which converge into one single peak at a magnetic field of 10 T (see Fig. 2(a)). The intensity of the peak maximum increases with increasing magnetic field to a value of about twice the maximum intensity at 0 T. However, the integrated intensity over both peaks remains almost constant. The merging into one single Bragg peak is an indication that the length of the spin cycloid increases continuously and finally diverges to infinity at a magnetic field of 10 T. The resulting spin structure is G-type antiferromagnetic, i.e. the spin cycloid collapses into a G-type antiferromagnetic spin arrangement. Upon decreasing magnetic field the peak splits again into two well separated peaks and the overall height of the peak maximum decreases with decreasing magnetic field see Fig. 2(b)). The initial lineshape, both in peak intensity and peak splitting has recovered when reaching 0 T again, i.e. the length of the spin cycloid reaches the same value of $65.7 \pm 1.0$ nm after the magnetic field cycle.

In order to confirm that only one magnetic Bragg peak remains at 10 T, RSMs were taken around the $\left(\frac{1}{2}\ \frac{1}{2}\ \frac{1}{2}\right)_{pc}$ and $\left(\frac{1}{2}\ \frac{1}{2}\ \frac{\bar{1}}{2}\right)_{pc}$ positions at 10 T and (see Figs. 3(a) and 3(b), respectively). The corresponding linescans along the $[11\bar{2}]_{pc}$ direction through the peak maxima are shown in Figs. 3(c) and 3(d). At zero magnetic field two magnetic Bragg peaks were present (see Figs. 1(a) and 1(c)) which indicates the presence of a spin cycloid in the BiFeO$_3$ thin film. At a magnetic field of 10 T the two magnetic Bragg peaks have converged into one single peak. The intensity of its peak maximum is almost twice the intensity of an initial single Bragg peak. Note, due to the orientation of the elliptical resolution function for the $\left(\frac{1}{2}\ \frac{1}{2}\ \frac{\bar{1}}{2}\right)_{pc}$ Bragg reflections, in Figs (b) and (d) only one single peak is expected even in the presence of a peak splitting.

Figure 4 shows the overall result of the magnetic field dependence of the neutron diffraction signal of the 100 nm thick BiFeO$_3$ thin film. The magnetic field dependence of the peak maximum is shown in Fig. 4(a) for an increasing (blue symbols) and a decreasing (red symbols) magnetic field. Note that the integrated intensity over both magnetic Bragg peaks remains almost unchanged. In Fig. 4(b) the magnetic field dependence of the length of the spin cycloid as extracted from the splitting of the two magnetic Bragg peaks, is presented. The distance between both Bragg peaks decreases continuously and converges towards zero at 10 T. This indicates that the length of the spin cycloid increases systematically and diverges to infinity at a magnetic field of about 10 T. The spin cycloid expands and finally collapses into a commensurate G-type antiferromagnetic spin structure. Since only one set of magnetic Bragg peaks was measured in this experiment, no conclusion can be drawn if the final antiferromagnetic structure is canted. It should be noted that both, the peak intensity and the length of the spin cycloid recover with decreasing magnetic field and exhibited a slight hysteresis behavior between 4 T and 10 T.

For bulk BiFeO$_3$ an abrupt phase transition was observed between about 16 T and 20 T. Based on measurements of the electric polarization, magnetization, and ESR in combination with theoretical modelling, the resulting phase was attributed to a canted antiferromagnetic structure [7,14,37-39]. In addition, an intermediate phase was observed by ESR and neutron diffraction experiments between about 10 T and 20 T [39,42]. The magnetic structure of this phase was described as conical. It is important to note that the neutron diffraction experiments on BiFeO$_3$ single crystals under high magnetic field did not indicate any change in the incommensurable splitting in both phases, i.e. the length of the spin cycloid did not change with magnetic field [41,42]. This is in contrast to our neutron diffraction experiments on a 100 nm $BiFeO_3$ thin film where the length of the spin cycloid increases systematically and diverges at 10 T. The present experiment on a $BiFeO_3$ thin film does not provide any information about the conical intermediate phase. With a maximum of 10 T this phase was perhaps not reached. In addition, a conical spin structure would cause two magnetic Bragg peaks in the direction perpendicular to the scattering plane. Due to the finite out-of-plane instrumental resolution, they would appear as a single magnetic Bragg peak in the centre. Since the incommensurate splitting of the two magnetic Bragg peaks did converge to zero, we would consider this scenario as unlikely.

An external magnetic field aligns the spins along the direction of the field. As the spin cycloid is partly anisotropic on a large scale, the applied magnetic field direction and magnetization direction are not same [15]. With increasing the magnetic field the Zeeman interaction strength can become comparable to the DM interaction. At a critical magnetic field this can lead to a suppression of the spin cycloidal. This can be expressed through the Landau-Ginzburg formalism of its free energy [25,38,39]. The Lifshitz invariant (a term in the Landau-Ginzburg theory that describes a coupling between the magnetization and its spatial variation due to the cycloidal spin structure) is increased by applying a magnetic field. For $BiFeO_3$ this can lead to the transformation of spin cycloid order into a simpler collinear antiferromagnetic state as observed in bulk $BiFeO_3$ [7,14,37-39,41.42]. Theoretical studies have suggested that a lower critical magnetic field is required to suppress the cycloidal spin order in epitaxially grown $BiFeO_3$ thin films [43,44]. This was confirmed by the Raman experiments of Agbelele *et al.* in $BiFeO_3$ thin films [40], where they reported a magnetic phase transition from a cycloidal spin structure to a G-type antiferromagnetic spin order in the range of $4\,T - 6\,T$ (depending on the chosen substrate) for $BiFeO_3$ thin films with a thickness of 70 nm grown on different (001)-oriented substrates, i.e. $DyScO_3$, $GdScO_3$, and $SmScO_3$. In comparison to our $BiFeO_3$ thin film grown on (110)-oriented $SrTiO_3$ substrate they had a significantly smaller epitaxial strain (see Refs. [30] and [40], respectively). Therefore, our critical value of 10 T for the collapse of the spin cycloid in a $BeFeO_3$ thin film is in accordance with the results of Agbelele *et al.* on differently strained $BiFeO_3$ thin films [40].

## 4. CONCLUSION

In summary, our neutron diffraction experiment using a triple-axis spectrometer revealed a magnetic field-dependent change of the spin cycloid in a 100 nm $BiFeO_3$ film deposited by PLD on a (110)-oriented $SrTiO_3$ substrate. The experimental result show a continuous change of the spin cycloidal upon applying a magnetic field. The length of the spin cycloid expands systematically and diverges to infinity at an applied magnetic field of 10 T where the spin cycloidal transforms into a commensurate G-type antiferromagnetic spin structure. A transition to a canted antiferromagnetic spin structure was observed before for polycrystalline $BiFeO_3$ and single crystals but at much higher magnetic fields in the range of 16 T to 20 T. In contrast, for bulk $BiFeO_3$ the incommensurable splitting of the magnetic Bragg peaks did not change with the magnetic field. Raman light scattering experiments showed that $BiFeO_3$ thin films with lower epitaxial strain when grown on different (001)-oriented substrates had a transition from the cycloidal phase to a collinear spin structure at a lower magnetic field of 4-6 T, depending on the substrate. Our neutron diffraction experiments directly proof that the length of the cycloid systematically extends and transitions continuously into a G-type antiferromagnetic structure. This provides important information about the electronic correlations and microscopic physical effects behind the origin of the spin cycloid in $BiFeO_3$. Furthermore, our results demonstrate that the application of a magnetic field offers a further route for the manipulation of the spin cycloid in multiferroics. Therefore, our experimental result will help to develop strategies to improve the linear magnetoelectric coupling in $BiFeO_3$ thin films with larger electric polarization, which is crucial for future technological applications.

## ACKNOWLEDGEMENT

We would like to thank ANSTO for providing the neutron scattering beamtime at the instrument TAIPAN through the proposals P7934 and P9961, and Guochu Deng for his support as beamline scientist.

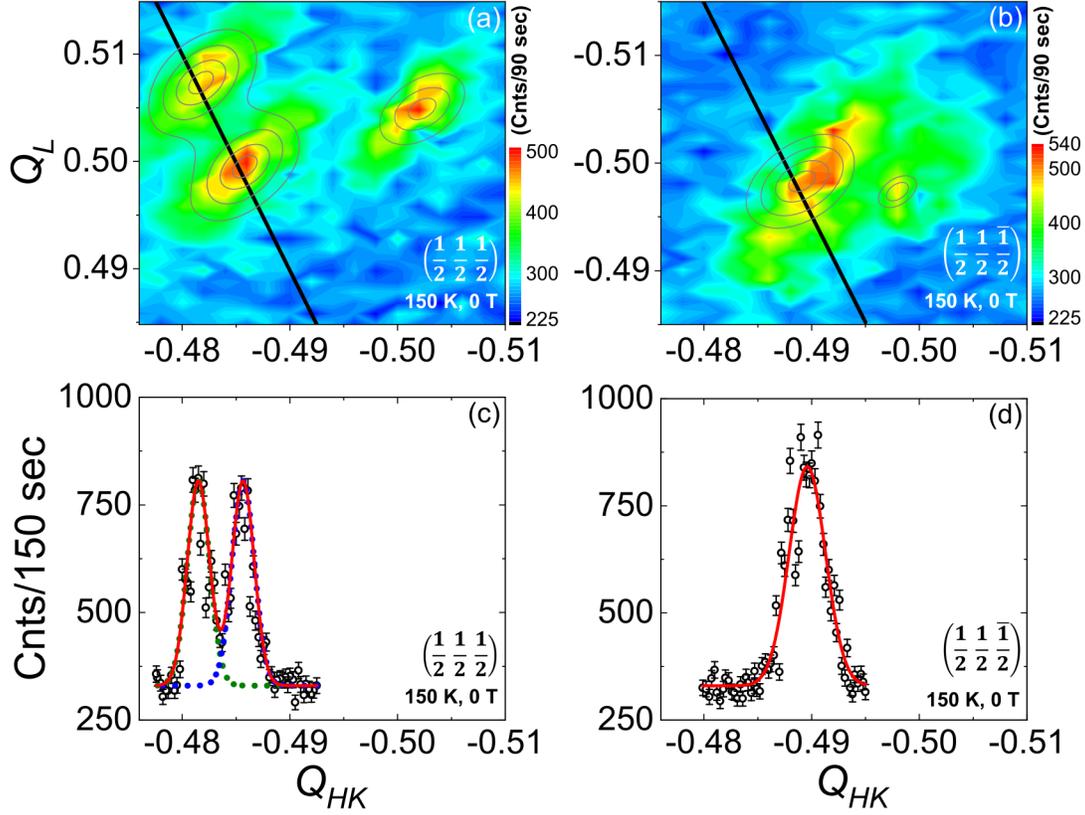

**Figure 1:** (a) and (c) Neutron scattering reciprocal space maps (RSMs) of the [110]-[001] plane of a 100 nm thick $BiFeO_3$ thin film grown on a (110)-oriented $SrTiO_3$ substrate. The data were taken at 150 K and 0 T. The Bragg reflection at $\left(\frac{1}{2}\,\frac{1}{2}\,\frac{1}{2}\right)_{pc}$ corresponds to second order contamination from the $SrTiO_3$ substrate. (a) RSM around the $\left(\frac{1}{2}\,\frac{1}{2}\,\frac{1}{2}\right)_{pc}$ Bragg reflection. Two additional Bragg peaks are observed. They originate from the spin cycloid of $BiFeO_3$. Their spacing, as highlighted by the black line, indicates the propagation direction, i.e. $[11\bar{2}]_{pc}$ and the length of the spin cycloid. (b) RSM around the $\left(\frac{1}{2}\,\frac{1}{2}\,\frac{\bar{1}}{2}\right)_{pc}$. Due to the elliptical instrumental resolution only one additional Bragg peak is visible. (c) and (d): corresponding line scans along the black lines in Figs. 1(a) and (b). The red lines are Gaussian lineshapes fitted to the experimental data with fixed linewidth and intensities for both magnetic Bragg reflections.

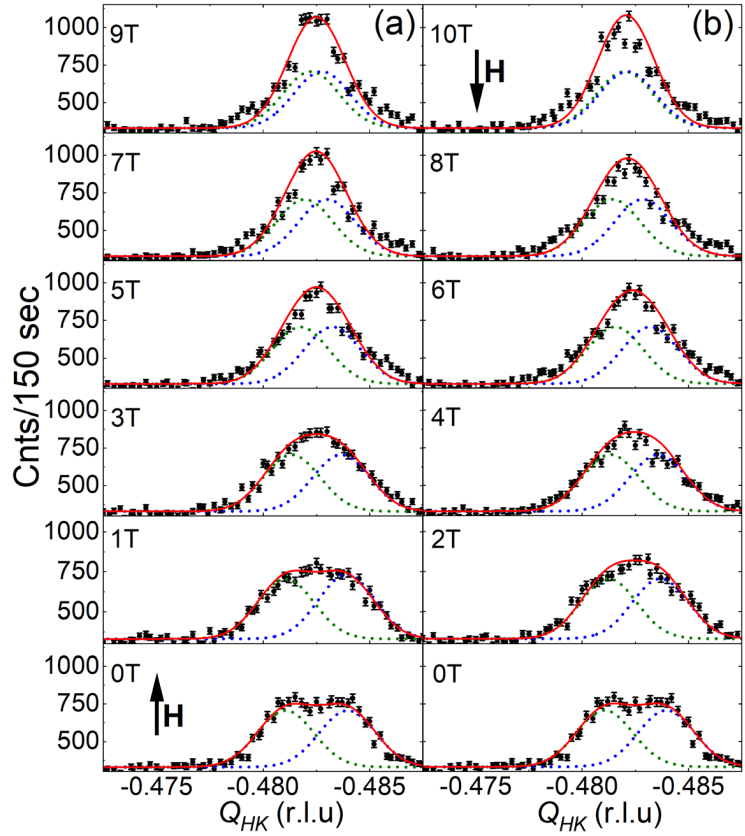

**Figure 2:** Neutron scattering data of the magnetic field dependence of the two incommensurate magnetic Bragg peaks in a 100 nm thick $BiFeO_3$ thin films. The linescans follow the black line shown in Fig. 1(a) and were measured at a temperature of 150 K in the magnetic field range from 0 T to 10 T, i.e. in (a) for an increasing magnetic field and in (b) for a decreasing magnetic field back to 0 T. The dashed and solid lines correspond to the result of a fit of two Gaussian lineshapes to the experimental data. For the fit the linewidth was kept constant at the value determined at zero magnetic field and the intensity of both Bragg peaks was kept constant at each magnetic field.

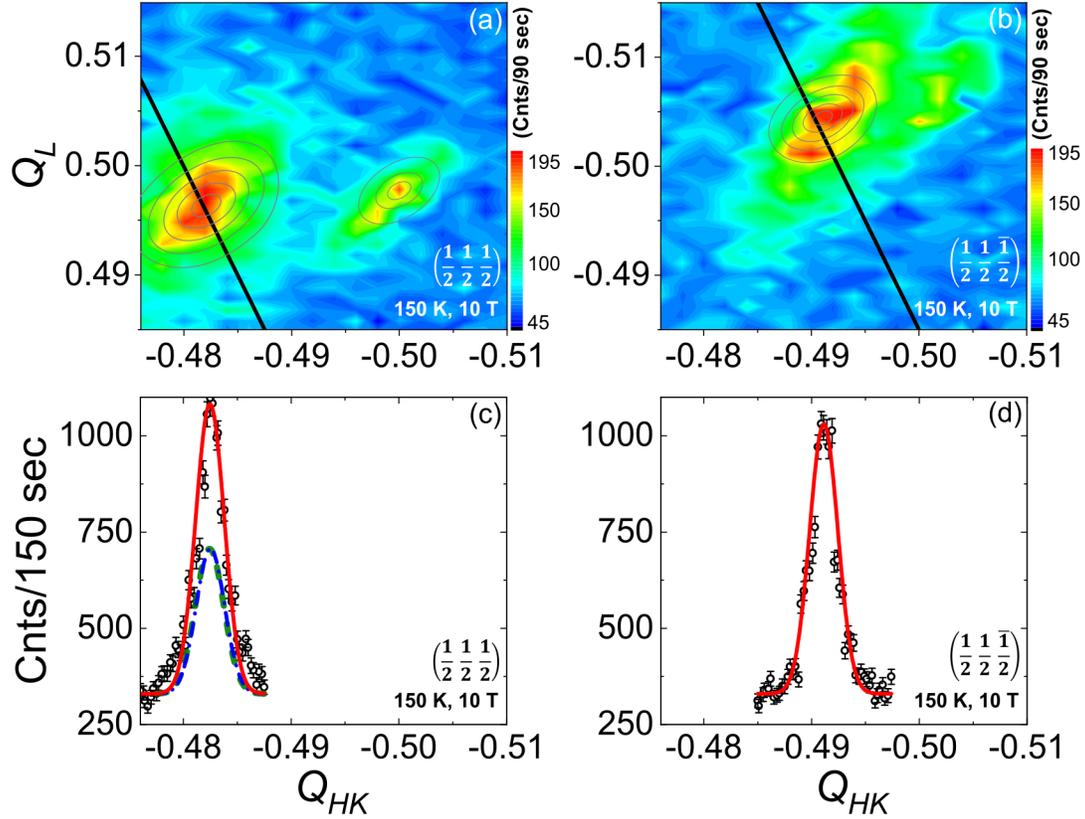

**Figure 3:** Neutron scattering data taken in the [110] x [001] scattering plane of the 100 nm thick BiFeO₃ thin film taken at a magnetic field of 10 T and a temperature of 150 K. Figure (a) and (b) show the RSMs around the $\left(\frac{1}{2}\,\frac{1}{2}\,\frac{1}{2}\right)_{pc}$ and $\left(\frac{1}{2}\,\frac{1}{2}\,\frac{\bar{1}}{2}\right)_{pc}$ Bragg reflections, respectively. The black lines represent the corresponding linescans in $[11\bar{2}]_{pc}$ direction as shown in Figs. (c) and (d), respectively. Gaussian lineshapes were fitted to the experimental data, as indicated by the dashed and solid lines.

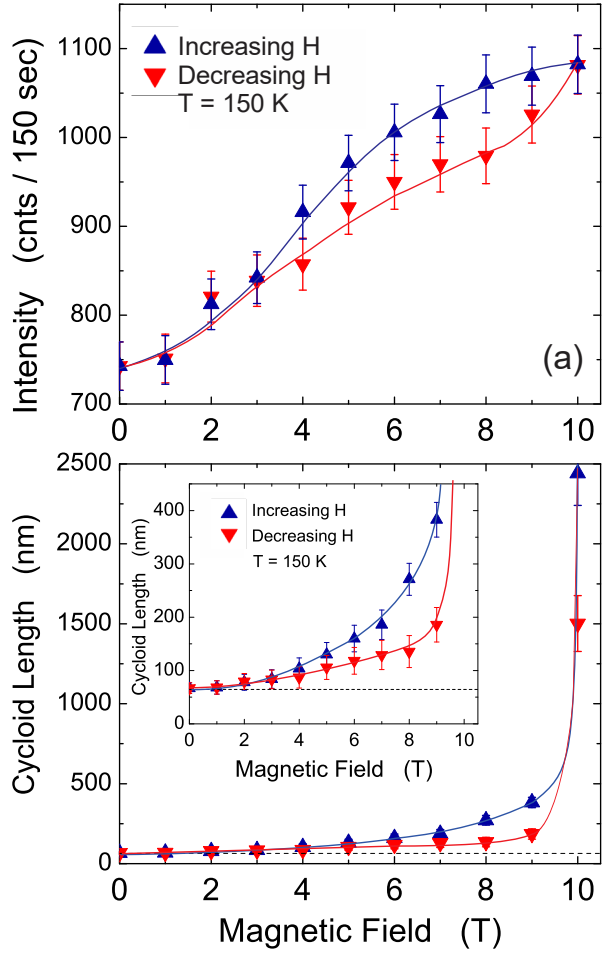

**Figure 4:** (a) Intensity of the pek maximum of the neuron scattering data as a function of magnetic field as shown in Fig. 2. The blue symbols indicate the data obtained for an increasing magnetic field and the red symbols present the data of a decreasing magnetic field. Note that the integrated intensity over both magnetic Bragg peaks remains almost unchanged. (b) length of the spin cycloid extracted from the splitting of the two magnetic Bragg peaks. The length of the spin cycloid is 65.7 ± 1.0 nm at zero magnetic field, expands systematically with increasing magnetic field and diverges towards infinity at a magnetic field of 10 T. The inset shows the zoomed-in data of the cycloidal length and the dashed horizontal line indicates the spin cycloid length of 63 nm obtained previously on bulk and thin film $BiFeO_3$ samples. The solid lines serve as a guides to the eye.